# A Data-Driven and Integrated Evaluation of Area-wide Impacts of Double Parking Using Macroscopic and Microscopic Models

Jingqin Gao[a], Kaan Ozbay[b], Michael Marsico[c]

[a] Department of Civil and Urban Engineering, New York University
[b] Department of Civil and Urban Engineering &Center for Urban Science and Progress, New York University
[c] New York City Department of Transportation

## 1. Introduction and Background

Double parking (DP) that often negatively affects traffic operations and safety is not a new phenomenon on urban streets. Evaluation of double parking impacts can be extremely challenging due to the difficulties in collecting data and modeling double parking behavior in a large-scale network. In New York City, clearly, there is no existing dataset that can provide an exact number of double parking activities. However, several other vast amounts of related temporal and spatial data are available citywide. In the light of all these complications, this study presents an enhanced data-driven framework that was originally proposed by the authors in a previous article *(1)* to integrate various data sources, such as parking violation tickets, 311 service requests and Twitter, to estimate the severity level of double parking in terms of its frequency and duration (Figure 1). The estimator uses Random Forests techniques and is proved to have 85% out-of-sample accuracy in estimating a combined index of double parking frequency and duration. The outputs from the estimator can then be feed into macroscopic or microscopic models as inputs with the goal of quantifying their impact on traffic.

The estimated double parking activities along with existing large scale models, such as Manhattan Traffic Model (MTM) that covers more than 800 intersections, creates opportunities to evaluate area-wide double parking impacts without other compounding factors. This allows local transportation agencies to identify not only the double-parking activity hotspots, but also the locations where transportation agencies can obtain the greatest benefits from eliminating double parking. Different parking enforcement and management strategies can be tested in the model to achieve more sustainable mobility in the study area.

On the other hand, macroscopic models should also be considered as an alternative approach to quantify double parking impacts in terms of lower complexity and computational speed compared with microscopic simulation models.

This paper first estimated the frequency and duration of double parking events using a well-known machine learning technique namely, Random Forests. Then area-wide double parking events are simulated in a large-scale microsimulation model to quantify their traffic impact. Next, we study the feasibility of using macroscopic queueing models that incorporate interruptions due to double parking instead of complex and computationally very demanding microsimulation models specially for large networks. Finally, the outputs from the both simulation and queueing models are used to identify double parking hotspots that need better enforcement or management.

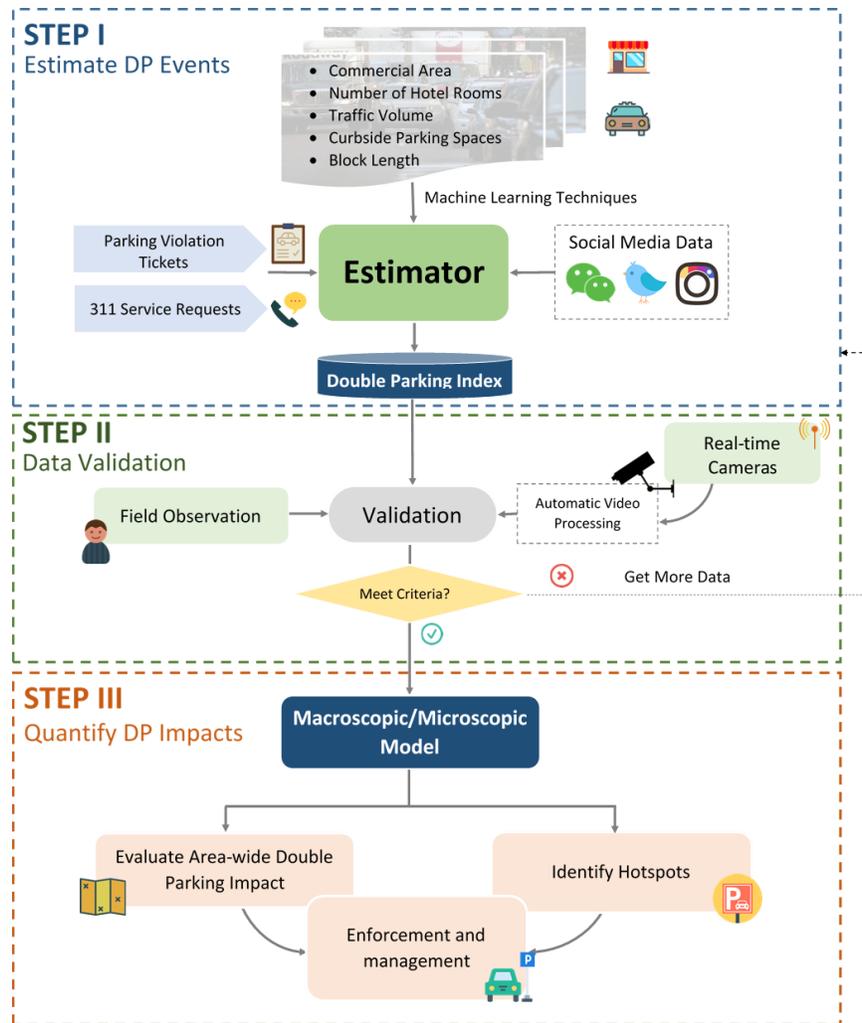

*Double Parking Index (DPI) is an indicator that represents the combined severity level of DP frequency and duration.
**Figure 1 Data-driven framework to quantify the impact of double parking**

## 2. Data Needs for Estimating Network-wide Double Parking Activities

Intuitively, double parking violation tickets and 311 service requests complaining about double parking blocking the traffic are directly related to the actual double parking occurrence, yet both of the datasets have their own selection and sampling biases. Supplemental data such as street characteristics (i.e. commercial area) and traffic information (i.e. traffic volume) are needed to address this issue. Table 1 shows the summary of datasets used in this study.

**Table 1 Description of Available Data**

| Data Name | Length | Source | Description |
|---|---|---|---|
| Parking Violations | 1 year | NYC OpenData* | Double parking related violation tickets for 2015 in NYC |
| 311 Service Requests | 6 years | NYC OpenData* | Double parking related requests from 2010 to 2015 in NYC |
| Twitter | 3 years | Twitter API | Geo-located tweets from 2013 to 2015 in Manhattan |
| Double parking Information | 1-3 days | Field Visit | Peak hour double parking occurrence and duration from field visits and real-time cameras. |
| Commercial Area | - | NYC OpenData* | Total commercial area for each street in study area |
| Hotel Rooms | - | Hotel Websites | Total number of hotel rooms for each street in study are |
| Traffic Volume | Varied | Field Visit | Average peak hour traffic volume for each street in study area |
| Curbside Parking Space | - | Field Visit | Number of curbside parking spaces during peak hour in study area |
| Block Length | - | NYC OpenData* | Block length for each street in study area |

*NYC OpenData *(2)* is the open data portal of New York City that offers access to government-produced data sets.

Besides traditional types of traffic datasets, the increasing use of crowdsourcing social media applications, such as Twitter *(3)*, has allowed novel approaches for obtaining extra information about human activities. This rich text information can allow us to better understand how people react to double parking activities. For example, people complaining about a double-parked vehicle blocking bike lanes or an ambulance having to double park because the curbside is occupied by trucks are both common occurrences in the twitter data. Emotional keywords can also provide indications of double parking severity and recurrent occurrence at certain locations.

## 3. Estimating Impacts of Double Parking
**Microscopic Model**

Manhattan Traffic Model *(4)* is a combined macroscopic/mesoscopic/microscopic simulation model of Manhattan with additional major arterials in the outer boroughs and New Jersey (Figure 2). While the macroscopic model disaggregates information from the regional New York Best Practices Model (NYBPM) and the mesoscopic model uses dynamic user-equilibrium following strict convergence criteria, the microscopic MTM model is a stochastic model based on dynamic traffic assignment that covers the street network from 14th Street to 66th Street, river-to-river. It is the largest, most comprehensive microsimulation model at the tri-state area.

In this study, a subarea in midtown Manhattan that contains 24 intersections and 36 roads are used. Double parking during the AM peak hour (8AM to 9AM) is coded as periodic section event that creates random incidents and is placed randomly throughout the street. During the time interval specified, the simulation creates single-lane incidents on any lane and position of the specified section following the time patterns defined *(5)*. Both the frequency and duration of the event can be either fixed or follow a selected distribution (i.e. Normal Distribution). The double-parking behavior in the model is calibrated (i.e. look ahead distance, passing speed) using the observation from multiple videos. Estimated number of Double Parking activities from our estimator are applied where no actual counts are available.

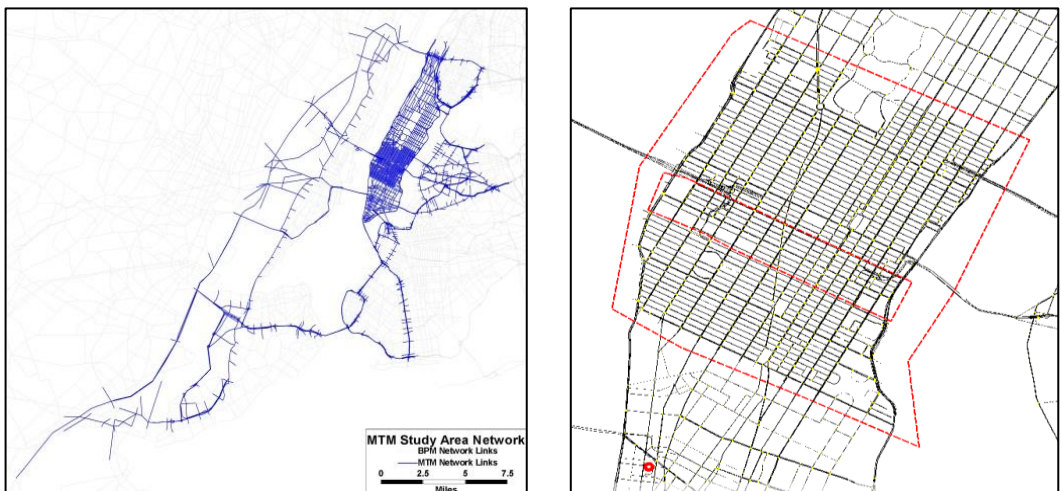

**Figure 2 MTM model and its primary micro-model area** *(4)*

**Macroscopic Model**
Although microscopic simulation models are effective for modeling this kind of extremely detailed behaviors, they are usually labor intensive to be developed and calibrated. As an alternative approach, macroscopic models can be magnitude of order inexpensive and faster to be developed and implemented. However, accuracy of such models need to be carefully validated. This study adopted Baykal-Gürsoy and Xiao's M/M/∞ model *(6-10)* with modification on vehicle type under double parking conditions. An M/M/∞ queueing model represents a system with a Markovian arrival rate, a Markov modulated service rates, and an infinite number of servers in the system *(7)*. The system in this study is designed with two server states "Failure (F)" or "Normal (N)" to illustrate the conditions with or without double parking. Denote E be the set of roadway links in the study area, $V=\{P,T\}$ to indicate the vehicle type, where *P* stands for passenger cars and *T* stands for commercial trucks. When the system experiences interruptions, the average number of vehicles on the roadway link *i* can be represented as follow (adopted from *(6)*):

$$N_i = \frac{D_i L_i}{v_i}\left[1 + \frac{\sum_j c_j f_{ij}\left(1-\frac{v'_i}{v_i}\right)}{\frac{1}{d_i}+\sum_j c_j f_{ij}}\left(1+\frac{\left(\sum_j c_j f_{ij}+\frac{v_i}{L_i}\right)(v_i-v'_i)}{\frac{v_i}{d_i}+\sum_j c_j f_{ij} v'_i+\frac{v_i-v'_i}{L_i}}\right)\right], \quad (i \in E, j \in V) \quad (1)$$

Subsequently, the average link travel time is:

$$t_i = \frac{L_i}{v_i}\left[1 + \frac{\sum_j c_j f_{ij}\left(1-\frac{v'_i}{v_i}\right)}{\frac{1}{d_i}+\sum_j c_j f_{ij}}\left(1+\frac{\left(\sum_j c_j f_{ij}+\frac{v_i}{L_i}\right)(v_i-v'_i)}{\frac{v_i}{d_i}+\sum_j c_j f_{ij} v'_i+\frac{v_i-v'_i}{L_i}}\right)\right], \quad (i \in E, j \in V) \quad (2)$$

Where $L_i$=length of link i (mile); $D_i$=hourly traffic demand on link i (veh/h); $f_{ij}$=frequency of double parking for vehicle type j on link i (events/hour), $d_i$=average duration time of double parking events on link i (h); v=average speed without double parking on link i (mph); and $v'_i$= average speed with double parking on link i (mph).

## 4. Preliminary Findings
The case study utilized a machine learning technique – Random Forests – to integrate different datasets from a variety of sources so that double parking events can be estimated at the street level. Instead of fitting the data into a model, machine learning techniques enable the computer to learn from the actual data. Random Forests, as one of the most popular and accurate machine learning methods that combine many decision tree predictors was used as a classifier to classify double parking events into a categorical response variable "Double Parking Index (DPI)" introduced by the authors in a previous study *(1)*. Figure 3 shows the estimation results for the study area.

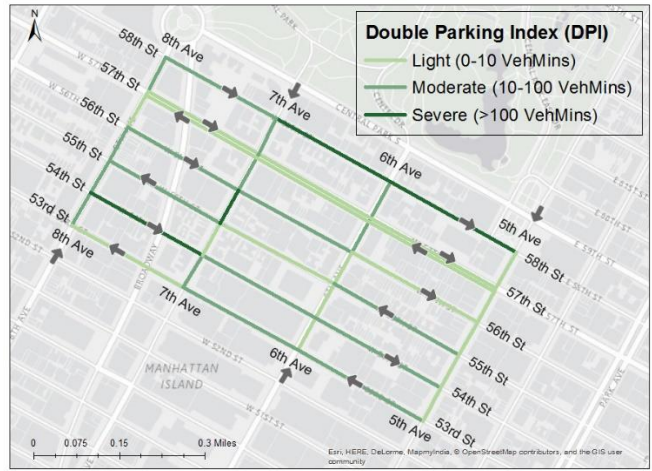

**Figure 3 Double Parking Index (DPI) Estimation Results**

The output generated by the random forests technique is then fed into a microsimulation model to evaluate area-wide double parking impacts. Figure 4 shows the preliminary results of the percentage difference in terms of average link travel time with or without double parking. Interestingly, although the 8$^{th}$ Avenue has a lower number of double parking events (3-11 events/link/hr) compared to the 6$^{th}$ Avenue (4-32 events/link/hr), it is more affected by double parking. One possible reason can be that majority of the double parking activities on the 6$^{th}$ Avenue are passenger cars or taxi drop-offs/pick-ups that have a relatively short duration less than one minute. However, the 8$^{th}$ Avenue has more commericial truck activities that have relatively longer durations, and all the traffic are heading to a roundabout (Columbus Circle) which can be usually a traffic bottleneck under saturated traffic conditions. Similar situations can be observed on the 5$^{th}$ Avenue and some crosstown streets.

The results suggest that priority should be given to the streets in dark red when considering parking enforcement as it may save more than 20% average travel time in the AM peak for these streets. Moreover, by eliminating all of the double parking activities in this subarea may reduce average travel time for each vehicle by 77.5 seconds during AM peak hour.

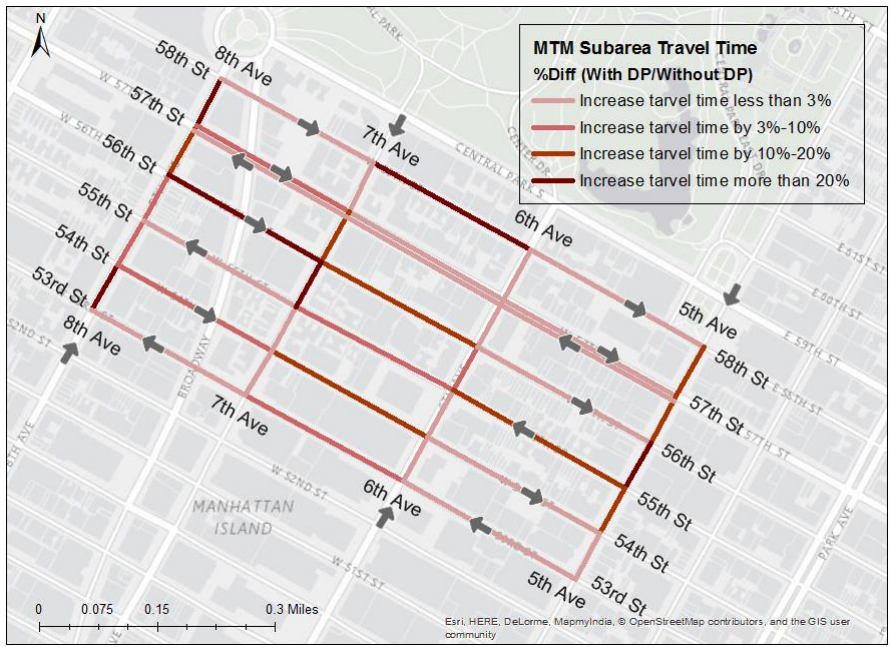

**Figure 4 MTM Subarea Results**

In our study, M/M/∞ queueing model is also found to be an effective approach and has the potential to be implemented instead of computationally demanding microsimulation models for large-scale network. Taking vehicle type into consideration enbles us to improve model performance. The full paper will cover more detailed computational results.

With our novel data-driven integrated framework for estimating the actual frequency of double parking, both microscopic or macroscopic models can be utilized to quantify area-wide impacts in the presence of double parking. The findings of this study can provide transportation agencies with useful insights on identifying locations that will experience the greatest benefits by removing problematic double parking. As a result, various parking enforcement and management strategies can be planned more effectively.